\documentclass[9pt,conference]{IEEEtran}
\IEEEoverridecommandlockouts
\usepackage{cite}
\usepackage{amsmath,amssymb,amsfonts}
\usepackage{graphicx}
\usepackage{xcolor}
\usepackage{diagbox}
\usepackage[a4paper, total={184mm,239mm}]{geometry}
\def\BibTeX{{\rm B\kern-.05em{\sc i\kern-.025em b}\kern-.08em
    T\kern-.1667em\lower.7ex\hbox{E}\kern-.125emX}}

\usepackage{comment}
\usepackage{subfiles}
\usepackage{url}
\usepackage{epsfig}
\usepackage{here}
\usepackage{multirow}
\usepackage{algorithm}
\usepackage{algorithmicx}
\usepackage{algpseudocode}
\usepackage[stable]{footmisc}

\renewcommand{\baselinestretch}{0.98}

\begin{document}

\title{QECOOL: On-Line Quantum Error Correction \\with a Superconducting Decoder for Surface Code}
%\begin{comment}

\author{\IEEEauthorblockN{Yosuke Ueno\textsuperscript{1}, Masaaki Kondo\textsuperscript{1,2}, Masamitsu Tanaka\textsuperscript{3}, Yasunari Suzuki\textsuperscript{4}, and Yutaka Tabuchi\textsuperscript{5, (a)}}
\IEEEauthorblockA{
\textsuperscript{1}Graduate School of Information Science and Technology, The University of Tokyo, Tokyo, Japan \\
\textsuperscript{2}RIKEN Center for Computational Science, Hyogo, Japan\\
\textsuperscript{3}Graduate School of Engineering, Nagoya University, Aichi, Japan \\
\textsuperscript{4}NTT Secure Platform Laboratories, Tokyo, Japan\\
\textsuperscript{5}Research Center for Advanced Science and Technology, The university of Tokyo, Tokyo, Japan}
\{ueno, kondo\}@hal.ipc.i.u-tokyo.ac.jp, masami\_t@nagoya-u.jp, yasunari.suzuki.gz@hco.ntt.co.jp, yutaka.tabuchi@riken.jp
}
%\end{comment}

\maketitle

\thispagestyle{empty}

\begin{abstract}
Due to the low error tolerance of a qubit, detecting and correcting errors on it is essential for fault-tolerant quantum computing. 
\textit{Surface code} (SC) associated with its decoding algorithm is one of the most promising quantum error correction (QEC) methods. 
% One of the challenges of QEC is its high complexity and computational demand.
QEC needs to be very power-efficient since the power budget is limited inside of a dilution refrigerator for superconducting qubits by which one of the most successful quantum computers (QCs) is built. 
In this paper, we propose an online-QEC algorithm and its hardware implementation with \textit{SFQ}-based superconducting digital circuits.
We design a key building block of the proposed hardware with an SFQ cell library and evaluate it by the SPICE-level simulation. 
Each logic element is composed of about 3000 Josephson junctions and power consumption is about 2.78~\mbox{\boldmath $\mu$}W when operating with 2~GHz clock frequency which meets the required decoding speed. 
Our decoder is simulated on a quantum error simulator for code distances 5 to 13 and achieves a 1.0\% accuracy threshold.
\end{abstract}
\renewcommand\thefootnote{\arabic{footnote})}	

\renewcommand\thefootnote{(\alph{footnote})}
\footnotetext[1]{Presently the author is with RIKEN Center for Emergent Matter Science.}
\renewcommand\thefootnote{\arabic{footnote})}	

\section{Introduction}
\label{sec:intro}

Quantum computers (QCs) are becoming an attractive computing paradigm, as the number of implementable qubits increases.
A QC with the largest number of qubits today has 53 qubits \cite{arute2019quantum} while real-world problems typically need more qubits. 
For example, a 256-bit RSA cipher with Shor's algorithm \cite{Shor_1997} requires about thousands of qubits. 
One of the challenges towards a rapid increase in the number of qubits is its high fragility of the quantum state due to decoherence and other noises, which necessitates error tolerance mechanisms.
A topological error-correcting code and associated quantum error correction (QEC) mechanism are widely studied to detect/correct errors \cite{Shor_physicalreview_1995}. 
Since it is not possible to directly observe qubits to detect errors, 
observational quantum bits called \textit{ancillary} qubits (\textit{ancilla bits}) are equipped in addition to informational qubits. 
Decoding an error-correcting code with ancilla bits to find erroneous qubits is a non-trivial task and requires a large amount of computation.

% with multiple physical qubits to represent a single logical qubit

%To increase the robustness of QCs to errors on qubits, huge amount of efforts has been paid for using multiple physical qubits to represent a single logical qubit and to construct error detection and correction mechanisms \cite{Shor_physicalreview_1995}.
%It is not possible to detect and correct errors by directly observing qubits because entangled quantum state is destroyed when they are measured. Therefore, researchers usually use methods to encode multiple physical qubits plus observational quantum bits called \textit{ancilla} bits in a indirect way to form a logical qubit and detect/correct errors of physical qubits by measuring ancilla bits. This process forms Quantum Error Correction (QEC). Decoding a logical qubit with ancilla bits to find erroneous qubits is a non-trivial task and requires large amount of computation. Thus, a classical computer system is typically used for this decoding process. 

One of the most promising QC implementations today is made up of superconducting qubits. 
To realize superconductivity and avoid thermal noise, they are operated in a cryogenic environment, especially at the effective temperature of a few millikelvin ranges.
While qubits are located on the millikelvin layer of a dilution refrigerator, most of the other components, including a processing unit for QEC, are located on a much higher temperature layer or even outside the refrigerator. 
This forces a QC to have many cables between different temperature layers, increasing the hardware complexity and latency of QEC processes which is one of the main hindrances of QC scalability. 
Performing QEC near the superconducting qubits dramatically alleviates this problem. 
However, the power budget of lower temperature layers in the dilution refrigerators is very restricted, for example, only tens of $\mu$W and around 1W in the millikelvin and 4-K layer, respectively. 
Extraordinary low-power QEC processing is necessary. 

%In order to improve the scalability of QCs with QEC, it is desirable to perform QEC in the same temperature environment as superconducting qubits. However, the permissible power consumption for mK layers of today's dilution refrigerators is only tens of $\mu$W. 
%Using CryoCMOS techniques\cite{cryoCMOS1993,cryoCMOS_for_QC} or even extraordinary low-power, superconductor digital circuits with single flux quantum (SFQ) is not easy to realize it. Hence, it is necessary to design an extremely power efficient QEC algorithm and its hardware implementation. 

One of the most feasible error-correcting codes for QEC is \textit{surface code} (SC) \cite{surface_code}. 
Since decoding the SC accurately is an NP-hard problem, approximating algorithms such as minimum-weight perfect matching (MWPM) \cite{MWPM_for_surface_code} is usually considered. As its computational complexity is still high, there have been proposed several low-cost decoders\cite{almost_linear,das2020scalable,holmes2020nisq}. 
Although correcting informational qubit errors is treated by a matching problem on a 2-D SC plane, we need to extend it to a 3-D SC lattice to deal with 
ancilla bit measurement errors whose error rate is equal to or even higher than that of informational qubits. 
Ancilla bits should be measured multiple times, and stacking each result temporally creates a 3-D SC lattice, which can be decoded by ordinary decoding algorithms for the 2-D plane with slight modifications. 
However, computational and circuit complexity greatly increases.

In this paper, we propose a power-efficient Quantum Error COrrection by On-Line decoding algorithm (QECOOL) for 3-D lattice and its hardware implementation using superconducting digital circuits with single flux quantum (SFQ). 
Most of the prior decoders for 3-D matching problems or direct extension of 2-D algorithms require a completed 3-D SC, which means the decoding process is performed after all the measurements are done for one QEC step. 
We refer to this as \textit{batch-QEC}. Instead of waiting for all the measurements, QECOOL starts a QEC process for every obtained SC plane with several consecutive accumulated SC planes. 
We call this \textit{online-QEC}. 
Since online-QEC can inherently achieve a shorter QEC cycle than batch-QEC, the number of qubit errors on a single decoding process is expected to be small, leading to higher error correction performance. 

% Our algorithm is based on a greedy minimum-weight perfect matching (MWPM) \cite{DRAKE2003211} for \textit{surface code} (SC) \cite{surface_code} which is one of the most feasible QEC codes. 
% Although most of the prior works for power-efficient decoder for QEC study a case for matching problems on 2-D structures, this problem setting requires the unrealistic assumption that ancilla qubits are measured without error.
% To adapt for measurement errors in SC, we need to build a 3-D lattice by repeating multiple measurements and stacking them temporally, and then solve matching problems on it.
% The prior works are not directly applicable to 3-D matching problems in a scalable way while our algorithm targets them.

% Furthermore, most of the prior decoders for 3-D matching problems assume that the decoding process is performed after all the measurement processes are completed, while the decoding process needs to be done after each measurement process in real quantum computer systems.
%We call the former ``batch-processing'', and the latter ``real-time processing'' among the above processing manners, and we focus on designing a ``real-time SC decoder''.

QECOOL is implemented with a specialized hardware logic with a spike signal-based matching architecture implemented by SFQ logic to realize the online-QEC with ultra-low-power and low-latency.
Since the semiconductor process technology of SFQ is not scalable as that of CMOS due to the necessity of constructing superconducting rings, a large number of memory blocks are not implementable. QECOOL adopts a distributed architecture approach and area-efficient design with small registers.

The contributions of this paper are summarized as follows:
\begin{itemize}
  \item We propose QECOOL, an on-line decoding algorithm for SC on a 3-D lattice that has significantly lower complexity than the standard MWPM and previously proposed decoders at the cost of slight degradation of error correction performance. 
  \item We design a power- and area-efficient SFQ decoder based on the above algorithm.
  \item We evaluate the power consumption and area of the design with a SPICE-based circuit simulator and show that our implementation can work at the 4-K layer with around 2,500 logical qubits.
  \item We evaluate the error correction performance of QECOOL with a quantum error simulator.
  \end{itemize}
The evaluation results show that QECOOL can perform sufficiently low-latency QEC while achieving very low-power consumption.

\section{Background and related work}
\subsection{Surface code\label{subsec:surface_code_intro}}

\begin{figure}[t]
    \centering
    \includegraphics[width=0.9\linewidth]{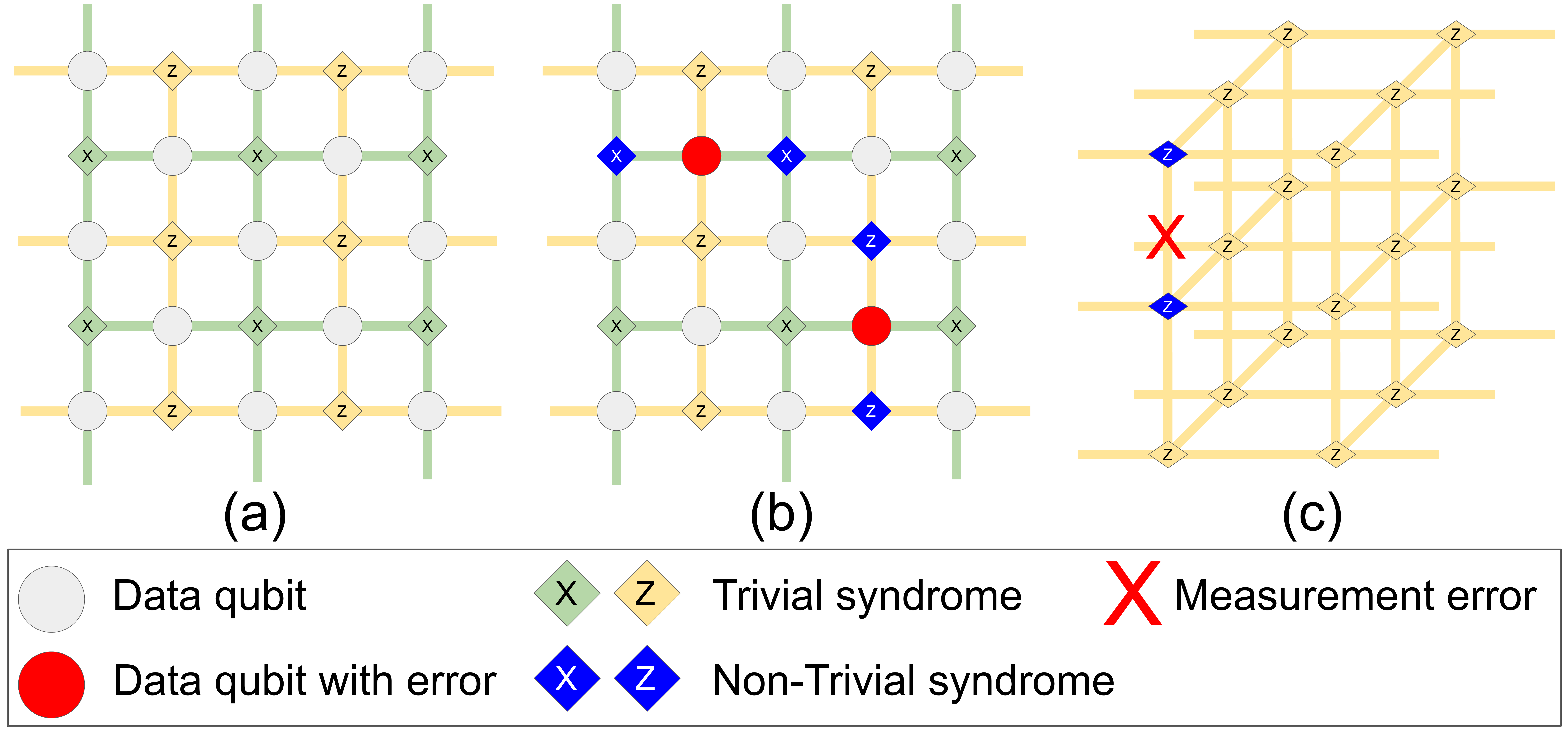}
    \vspace{-3mm}
    \caption{Graphical illustration of surface code}
    \label{fig:surface_code}
    \vspace{-3mm}
\end{figure}

Surface code (SC) is a quantum error correction code that provides highly reliable QEC. 
Figure~\ref{fig:surface_code} shows the schematic view of a SC for code distance $d = 3$.
It is implemented on a 2-D grid array of two types of physical qubits, data and ancilla qubits, which are represented as circles and squares in Fig.~\ref{fig:surface_code} (a), respectively.
Data qubits are redundantly used to represent a single logical qubit. 
Ancilla qubits interact with their neighboring data qubits, and their measurements form the \textit{error syndrome} (Fig.~\ref{fig:surface_code} (b)).

The purpose of decoding SCs is to find the types and location of qubit errors from error syndromes. 
There are two types of ancillary qubits, $X$-and $Z$-ancilla qubits, corresponding to Pauli-$X$ (bit-flip) and Pauli-$Z$ (phase-flip) error, respectively. 
Another error type, Pauli-$Y$ error, is represented by combinations of $X$ and $Z$ errors. 
Each ancilla bit works as the parity of the number of errors on its four neighbors.
Since ancilla qubits for $X$ or $Z$ errors are alternatively structured, decoding of SC can be independently performed for the $X$ and $Z$-syndrome. 

Decoding the SC is a process to find the most likely error correction operation from error syndromes. 
%The most desirable property of the decoder is that it can perform the recovery operation with the lowest logical error rate.
Since finding such an operation is NP-hard, it is common to reduce the problem's complexity to a polynomial-time solution by approximating a part of the problem. 
The most promising approach is applying the minimum-weight perfect matching (MWPM) problem\cite{surface_code}.
A variety of decoders using machine learning\cite{varsamopoulos2019comparing}, or union-find\cite{almost_linear} have also been proposed. 

Given that an ancilla bit measurement is susceptible to read errors, QEC needs multiple ancilla bit measurements over a series of quantum operations. 
This procedure creates multiple 2-D SC snapshots. 
By stacking them in the vertical direction, creating a 3-D lattice-shaped SC\cite{topological_memory} as shown in Fig.~\ref{fig:surface_code} (c), error syndromes are detected by solving the matching problem on the 3-D lattice even with measurement errors.

\subsection{Single Flux Quantum logic \label{subsec:SFQ_intro}}

The single flux quantum (SFQ) logic is a digital circuit that uses superconductor devices.
Information processing in SFQ is performed with magnetic flux quanta stored in superconductor rings with Josephson junctions (JJs) that act as switching devices.
The presence and absence of a single magnetic flux quantum ($\Phi_0 = 2.068\times10^{-15}$~Wb) represent logical `1' and `0', respectively. 
% An impulse shaped voltage pulse, called an SFQ pulse, is generated only when an SFQ travels across a JJ.
Several studies have shown SFQ-based microprocessor designs and demonstrated their ultra-high-speed and low-power operations \cite{Sato_ieeeTr_2017}. However, more power efficiency is required to work with superconducting qubits in the same temperature layer. 

%Some researches have been succeeded in demonstrations of an 8-bit microprocessors composed of more than 10,000 JJs which bit-serial operates at 50~GHz\cite{Sato_ieeeTr_2017} 
% and a gate-level-pipelined, bit-parallel multiplier up to 48~GHz with power consumption of 5.6~mW\cite{48GHzSFQ,2019nagaoka_isec}.

%Since SFQ circuits consist of superconductor devices, they work only in a cryogenic environment with around~4 Kelvin.
%Because of this constraint, finding the appropriate applications for SFQ is not a trivial task.
%While research on quantum computers has been actively developed, solid-state qubits based on superconductor devices is promising for
%quantum computers, which operate in a cryogenic environment.
%In this work, we apply SFQ circuits to quantum computers, especially rapid and efficient quantum error correction.

\subsection{Related Work on Hardware-Efficient Decoder}
\label{subsec:related_work}
A union-find based surface code decoder is recently proposed and has attracted much attention because of its accuracy and simplicity\cite{almost_linear}.
A practical architecture for the union-find decoder is first proposed by Das et al.\cite{das2020scalable}.
In their work, the fully pipelined hardware implementation contributed to the higher speed of the decoding process.

Using cryogenic computing as peripherals of QCs to perform controls, including QEC, has been actively studied\cite{holmes2020nisq,2017micro_qureshi,qureshi_2017ACM}.
Holmes et al.~\cite{holmes2020nisq} designed a novel algorithm with SFQ to find quantum errors online for near-term quantum systems. In their work, multiple pairs of two flipped ancilla bits can be found in parallel to reduce the latency of the decoding process. 
They implemented a mechanism that makes ``agreement'' among ancilla bits to find pairs.
% This process is not simple and requires a relatively complex circuit design.
Our base decoder algorithm is inspired by their mechanism but differs in the fact that QECOOL does not need the agreement mechanism to simplify the design. Moreover, none of the prior work addresses online-QEC on 3-D lattice for faster QEC cycle.

\section{Spike-based On-line Quantum Error Correction\label{sec:neuromorphic_algorithm}}

\subsection{Base decoding algorithm \label{subsec:summary_of_algorithm}}

We first describe how to decode a SC on 2-D structures in QECOOL, which is inspired by the greedy algorithm of minimum-weight perfect matching problems\cite{DRAKE2003211}. The purpose of the decoder is to find pairs of erroneous ancilla bit locations such that the total Manhattan distance of all the pairs is minimized. 
We introduce a module named \textit{Unit} associated with each ancilla bit. 
Each Unit is connected to neighboring four Units forming a 2-D grid structure as same as $X$ or $Z$ stabilizers.
We also introduce a \textit{Controller} module to orchestrate all the Units in a logical qubit.

\algrenewcommand\algorithmicindent{0.4em}
\begin{algorithm}[t]
\label{alg:decoder}
 \footnotesize
 \caption{Spike-based on-line QEC for 3-D Surface code}
  \begin{tabular}{cc}
\hspace{-4mm}
\begin{minipage}[t]{.22\textwidth}
 
%%% old code (for batch processing)
% \begin{algorithmic}[1]
% \State {\bf MeasureEachUnit:}
% \For{$t = 0$ to $N_{depth}$}
%    \State A = checkAncilla()
%    \If{$t == 0$}
%    \State Reg[0] = A
%    \Else
%    \State Reg[$t$] = Reg[$t$-1] $\oplus$ A
%    \EndIf
%    \State Sleep(Mcycle)
% \EndFor
% \end{algorithmic}

 \begin{algorithmic}[1]
 \State {\bf MeasureEachUnit:}
 \State {$m = 0$}
 \While{true}
    \State A = checkAncilla()
    \If{$m == 0$}
    \State Reg[0] = A
    \Else
    \State Reg[$m$] = Reg[$m$-1] $\oplus$ A
    \EndIf
    \State{$m = m + 1$}
    \State Sleep(Mcycle)
 \EndWhile
 \end{algorithmic}

%%% old code (for batch processing)
%\vspace{0.5em}
%\begin{algorithmic}[1]
% \State {\bf Controller:}
%\For{$C = 1$ to $N_{limit}$}
%\For{$b = 0$ to $N_{depth}$}
% \For{$i = 0$ to $N_{row}$}
%    \State currentRow = $i$
%    \For{$j = 0$ to $N_{col}$}
%       \State giveToken(i,j)
%       \State RestartUnit(b)
%       \State {\bf while} !getFinish() \&\& \\~~~~~~~~~~~~~~~~~~~~~~~!Timeout()
%    \EndFor
% \EndFor
% \State sendResetFlag()
%\EndFor
%\EndFor
%\end{algorithmic}

\vspace{0.5em}
\begin{algorithmic}[1]
 \State {\bf Controller:}
 \State{start\_loop:}
\For{$C = 1$ to $N_{limit}$}
\For{$b = 0$ to $N_{depth}$}
 \State{shift = true}
 \For{$i = 0$ to $N_{row}$}
    \State currentRow = $i$
    \For{$j = 0$ to $N_{col}$}
       \If{$m-b > th_{v}$}
           \State giveToken(i,j)
           \State RestartUnit(b)
           \State {\bf while} !getFinish() \&\& \\~~~~~~~~~~~~~~~~~~~~~~~!Timeout()
        \EndIf
       \State{shift\&=!Unit(i,j).Reg[0]}
    \EndFor
 \EndFor
 \State sendResetFlag()
 \If{shift}
 \State{SHIFTREG()}
 \State{goto start\_loop}
 \EndIf
\EndFor
\EndFor
\end{algorithmic}

\vspace{0.5em}
\begin{algorithmic}[1]
\Procedure{Spike}{row, flag}
\If{row == currentRow}
  \If{flag == 1}
     \State sendSpikeEast();
  \Else
     \State sendSpikeWest();
  \EndIf
\Else
  \If{flag == 1}
     \State sendSpikeSouth();
  \Else
     \State sendSpikeNorth();
  \EndIf
\EndIf
\EndProcedure
\end{algorithmic}
\end{minipage}

\begin{minipage}[t]{.35\textwidth}
\begin{algorithmic}[1]
\Procedure{ShiftReg}{}
\If{$m > 0$}
     \For{$i = 0$ to $N_{depth}-2$}
        \State Reg[i] = Reg[i+1]
     \EndFor
    \State{$m = m - 1$}
\EndIf
\EndProcedure
\end{algorithmic}
\vspace{0.5em}

\begin{algorithmic}[1]
 \State {\bf RestartUnit(Input: b)}
 \If{Token == 1}
   \State FlagToken = 1
   \If{Reg[$b$] == 1}
     \State requestSpike()
     \For{$t = b$ to $N_{depth}$}
       \If{(S = getSpike()) != NULL}
       \State Dir = rotate(S)
       \State correctQubit(Dir)
       \State sendSyndrome(Dir)
     \EndIf
     \If{$t$ != b \&\& Reg[$t$] == 1}
        \State sendController("Finish")
     \EndIf
     \EndFor
   \Else
     \State sendController("Finish")
   \EndIf
 \Else 
   \For{$t = b$ to $N_{depth}$}
   \If{Reg[$t$] == 1}
     \State SPIKE(self.row,FlagToken)
     \If{getCorrect()}
       \State Reg[$t$] = 0
       \State sendController("Finish")
     \EndIf
  \Else
    \If{(S = getSpike()) != NULL}
      \State Dir = rotate(S)
      \State SPIKE(self.row,FlagToken)
      \If{getCorrect()}
        \State correctQubit(Dir)
        \State sendSyndrome(Dir)
      \EndIf
    \EndIf
 \EndIf
 \EndFor
 \EndIf
 \end{algorithmic} 
\end{minipage}  
  \end{tabular}
\end{algorithm}

QECOOL algorithm is shown in Algorithm~1.
Though the algorithm shows online-QEC on a 3-D lattice, it can be considered as an algorithm for a 2-D plane if $N_{depth}$, the number of depth in the vertical direction (V-direction), and $th_{v}$, the maximum length or threshold in the V-direction to search for matchings, are set to 1 and -1, respectively. 
Each Unit has a unique row and column IDs originated by the Unit at the top-left corner.
When each ancilla bit is measured, its value (0 and 1 indicate consistent and inconsistent for connected qubits, respectively) is stored into \textit{Reg} on the associated Unit (MeasureEachUnit in Algorithm~1). 
In a QEC phase, the Controller assigns a Token to each Unit from the top-left (northwestern) corner. 
The decoding process shown in the ``RestartUnit'' procedure of Algorithm 1 is summarized as follows. 
Note that this procedure is supposed to be done in parallel in all the Units.

\begin{figure}[t]
    \centering
    \includegraphics[width=\linewidth]{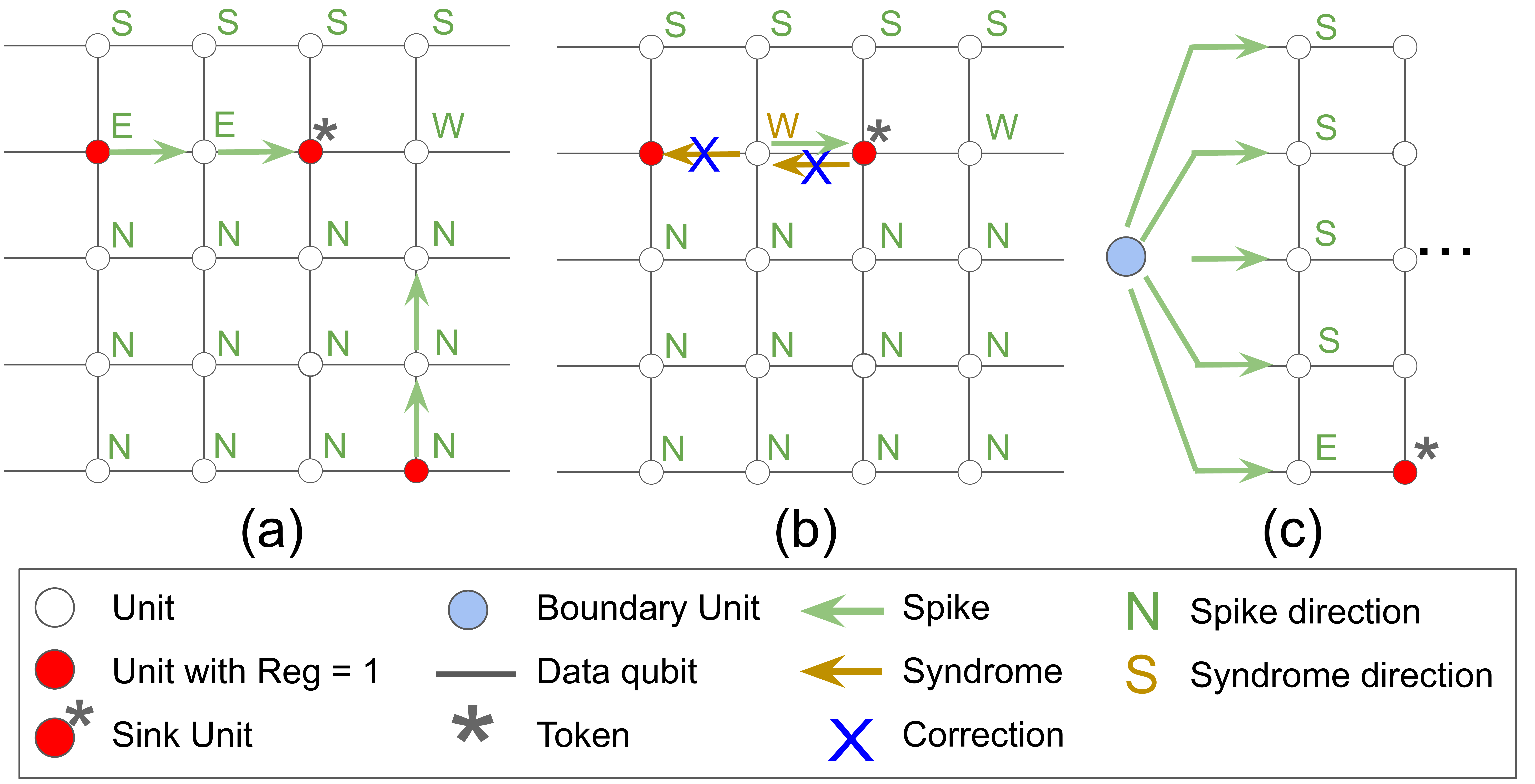}
    \caption{Overview of the processing of our algorithm.}
    \vspace{-2mm}
    \label{fig:algorithm}
\end{figure}

\begin{enumerate}
    \item If a Unit gets the Token, it checks its own Reg. If it is 1, the Unit becomes the \textit{sink} Unit and requests all the other Units to send a Spike toward the sink if their Reg is also 1, and then waits for the first Spike to come. If not, the Token is passed to the next Unit through the Controller.
    
    \item When the Spike request comes to a non-sink Unit, it checks its own Reg. If Reg is 1,
    it initiates a Spike toward the sink Unit (Fig.~\ref{fig:algorithm}(a)). The direction to send is determined locally by comparing its row ID and \textit{currentRow} set by the Controller, and also \textit{FlagToken} which indicates whether the Token is already passed in the past (see SPIKE procedure in Algorithm 1).
    
    \item For a non-sink Unit with Reg value is 0, it may get a Spike from one of its neighboring Units. In this case, it passes the Spike to one of the other neighboring Units so that the Spike reaches the sink Unit. The direction to pass is determined in the same way as Step 2). It also calculates and saves the relative direction of the coming Spike by rotating 180 degrees of the incoming port's direction. This direction is used to generate an error syndrome in the following Step 4).

    \item If the first Spike comes to the sink Unit, it generates a correct signal to an informational qubit on the coming Spike direction. It also sends \textit{Syndrome} signal to that direction to correct errors of qubits on the error syndrome (Fig.~\ref{fig:algorithm}(b)). The Syndrome signal is passed toward a Unit that initiates the Spike using the saved direction in Step 3) (corresponding to \textit{correctQubit()} and \textit{sendCorrect()} in Algorithm 1).
\end{enumerate}
This algorithm ensures to find the closest Unit pairs whose Reg values are both 1 by sending a Syndrome signal to the initiator of the first coming Spike. However, this does not necessarily find a set of pairs that minimize the total distance since the sink node is sequentially allocated to Units. Therefore, we limit the maximum number of hops to propagate a Spike in a single run of the above steps and increase it iteratively. 
The \textit{Controller} procedure in Algorithm 1 implements this by setting a timeout after giving the Token to a Unit.

If an error syndrome appears from one of the Units in the SC toward the outside (Fig.~\ref{fig:algorithm}(c)), there is no proper matching pair between erroneous ancilla bits. To deal with such a case, we introduce additional Units called \textit{Boundary Units} at the edges of the SC. The Boundary Units never get a Token from the Controller, but always send a Spike upon a \textit{requestSpike()} call\footnote{To prioritize matching between normal Units, the Spike timing of Boundary Units is adjusted in the implementation.}.

To deal with matching problems on a 3-D lattice, results of multiple rounds of measurement for an ancilla bit are stored in each Unit instead of actually having Units for every 3-D lattice points. 
\textit{Reg} is extended to be an array to hold several measurement results.
For every new measurement, the value of the ancilla bit is XORed with the latest value stored in Reg (MeasureEachUnit process in Algorithm 1). 
% This procedure is to indicate that a possible measurement error happens in the measurement. 

% \begin{figure}[t]
%     \centering
    % \includegraphics[scale=0.14]{img/boundary_dummy_by_controller.pdf}
    % \caption{logical qubit for the coding distance $d = 3$ with boundary dummy unit}
    % \label{fig:virtual_bit}
% \end{figure}

\subsection{Online-QEC for faster error correction cycle\label{subsec:3d-surface}}
\begin{figure}[t]
    \centering
   \includegraphics[width=\linewidth]{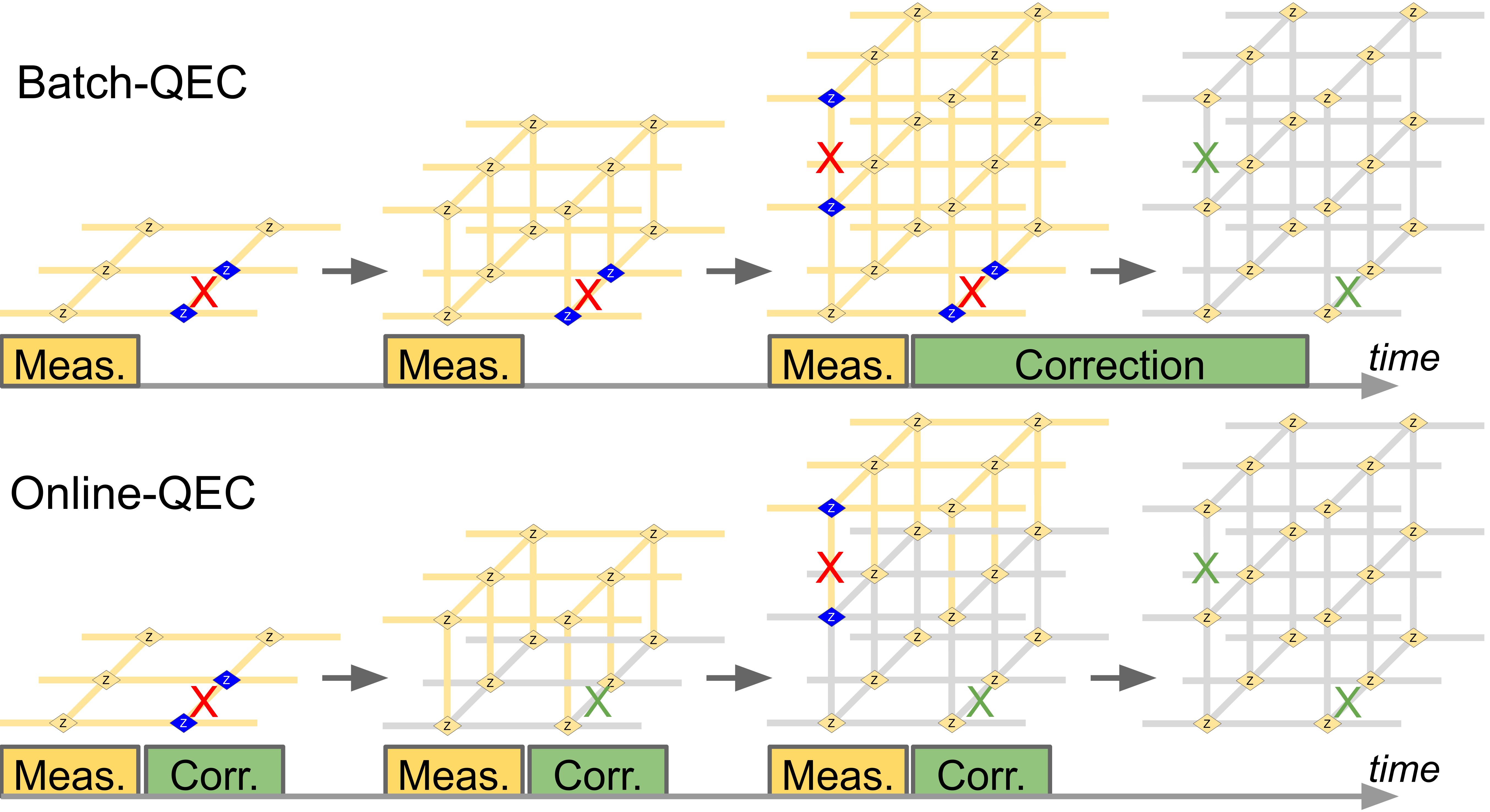}
   \caption{Concept of Batch- and Online-QEC}
   \vspace{-3mm}
   \label{fig:batch_and_online}
\end{figure}

Figure~\ref{fig:batch_and_online} compares the concepts of batch and online-QEC for a distance-3 SC with measurement errors. In batch-QEC, an error correction phase can be scheduled after three measurements are processed. On the contrary, an error correction phase is associated with each measurement step in online-QEC. This helps correct qubit errors as soon as they occur, which reduces the number of qubit errors in a single QEC phase and makes the matching process easy. However, QEC with only a single SC plane should be avoided due to the occurrence of ancilla bit measurement errors.
Hence, the Controller waits for several measurements have done before starting a QEC phase. This is ensured by 
having at least $th_{v}$ measurement results in Reg for each Unit (L.9 in \textit{Controller} code in Algorithm 1). 
The threshold value of $th_{v}$ controls a trade-off between the error-to-correction latency and error correction performance. We evaluate the appropriate $th_{v}$ in the next Subsection.

%This algorithm is assumed to be for online-QEC, which means the measurement (MeasureEachUnit) and the error correction process (Controller and RestartUnit) are executed simultaneously.
%However, if arbitrary length matchings are searched without a sufficient number of measurements stored in Reg, the algorithm will not be able to deal with measurement errors.
%Hence, the Controller needs to wait for several measurement processes to finish if Reg stores less than $th_{v}$ newer measurements than the base depth (L.9 in Controller).
%$th_{v}$ is responsible for the trade-off relationship between the latency and error correction performance, and the appropriate value of $th_{v}$ is determined in the next subsection.

In the Spike generation, multiple bits in a Reg are sought step-by-step from the oldest to the newest measurement. 
Once a non-sink Unit finds the value of 1 in the current bit position, it sends a Spike to the sink Unit. 
This process is repeated by changing the \textit{base depth} which points to the start position of the Reg for seeking. Otherwise, the matching process is the same as the 2-D case. 

\subsection{Necessary vertical depth for online-QEC \label{subsec:pre_error_correction_performance}}
%In this subsection, we evaluate error correction performance of the proposed circuit.

\begin{figure}[t]
    \centering
    \includegraphics[width=\linewidth]{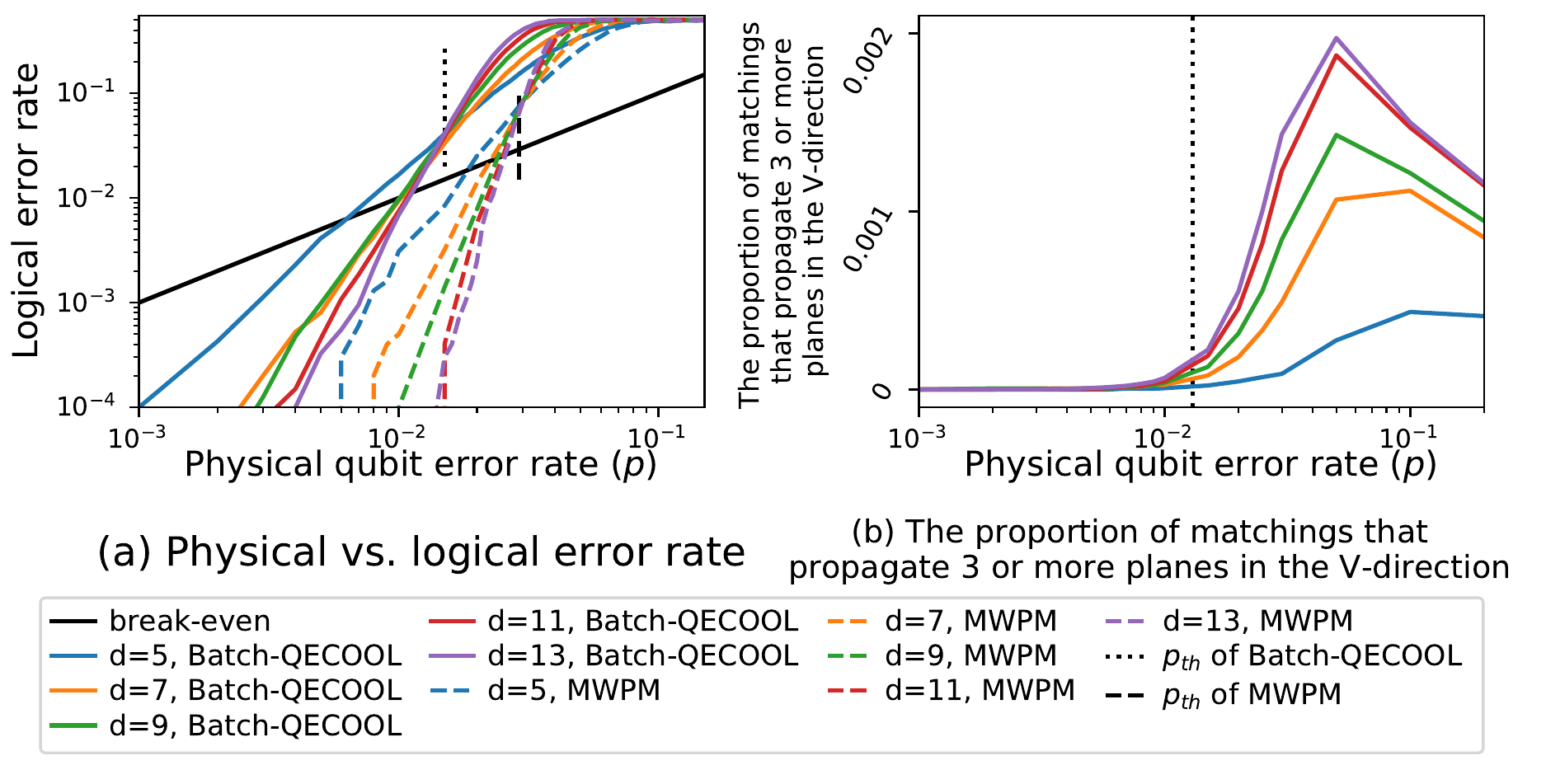}
    %\vspace{-4mm}
    \caption{(a) Error rate scaling for the MWPM decoder and batch-QECOOL. We can observe the threshold value of QECOOL at approximately 1.5\% physical error rate. (b) The proportion of matchings that propagate through three or more planes in the vertical direction.}
    \vspace{-3mm}
    \label{fig:error_correction_performance}
\end{figure}

%As described in Section~\ref{sec:intro}, our research aims to design a real-time processing algorithm and its hardware implementation.
%We evaluate error correction performance of the QECOOL algorithm when executed in a batch-QEC manner.

We evaluate how the vertical depth threshold $th_{v}$ affects the error correction performance of QECOOL.
We perform a spike-based QEC shown in Algorithm 1 with the batch-QEC manner by setting $N_{depth}$ and $th_{v}$ to $d$ and $-1$, respectively, indicating that the Controller process is executed after $d$ times measurement. 
We call this method as \textit{batch-QECOOL}. 

%Based on the preliminary evaluation results, we can determine the appropriate $th_{v}$ and estimate the resources required for the QECOOL decoder. 

The error-correcting performance is evaluated numerically by a quantum error simulator. 
We simulate the QECOOL algorithm with both informational and ancilla qubit errors using the phenomenological noise model\cite{topological_memory}. 
We show logical $X$ error rates versus physical Pauli-$X$ error rate as the performance indicator\footnote{Under the assumption that Pauli $X$, $Y$, and $Z$ errors occur stochastically, $Y$ errors can be considered as a simultaneous $X$ and $Z$ error.
Therefore, even if $X$ and $Z$ errors are corrected independently, all errors can be decoded correctly. Thus, we show only the case of $X$ error.}. 
We assume the error probabilities of data and ancilla qubits are equal.
Note that the same experimental setup is used in Section~\ref{subsec:error_correction_performance}.

Figure~\ref{fig:error_correction_performance}(a) shows the log scale plots of physical qubit error rate and logical-$X$ error rate for Pauli-$X$ errors. 
The solid black line is the break-even (physical error rate = logical error rate) value. 
The other solid lines show the results for batch-QECOOL, whereas the dashed lines represent cases for MWPM\cite{MWPM_for_surface_code}. 

We use \textit{threshold values} to evaluate the error correction performance of our algorithm.
The threshold value, $p_{th}$, is defined for each decoding algorithm and represents the value of physical error rate $p$ that satisfies the following properties; if the physical error rate $p$ is less than $p_{th}$, the logical error rate $p_{L}$ decreases as the code distance $d$ increases.
It can be defined as the value of $p$ at the intersection of $p$ and $p_{L}$ plotted for several code distances $d$.
The higher the $p_{th}$, the better the decoding algorithm.
$p_{th}$ of batch-QECOOL can be obtained from Fig.~\ref{fig:error_correction_performance}(a) and it is around $p = 0.015$, while that of MWPM is 0.03.

Figure~\ref{fig:error_correction_performance}(b) shows the proportion of matchings that propagate through three or more planes in the vertical (temporal) direction on a 3-D lattice. While three or more propagation in the vertical direction 
happens many times in larger physical error rate $p$, it is negligible for relatively smaller $p$, especially when $p$ is less than $p_{th}$. In general, QEC decoders require that the physical error rate $p$ is smaller than their $p_{th}$ for realistic error correction. This indicates having three SC planes (or Reg of three entries in QECOOL) for online-QEC is almost satisfactory. Thus, we assume $th_{v} = 3$ in the following sections. 

%This figure suggests that our algorithm have a low ability to find the matchings which contains many V-direction paths even if the coding distance $d$ is larger because of its greedy nature.
%In general, QEC decoders require that the physical error rate $p$ is quite smaller than their $p_{th}$ for efficient operation.
%For example, assuming $p$ to be $10^{-3}$, one-tenth of the $p_{th}$ of batch-QECOOL, the proportion of matchings with 3 or more V-paths is 0 or a negligible value, regardless of $d$.
%These results suggest the required Reg size in hardware implementation.

\section{Hardware Implementation}
\label{sec:hardware_section}

\subsection{Hardware architecture}

\begin{figure}[t]
    \centering
    \includegraphics[width=0.85\linewidth]{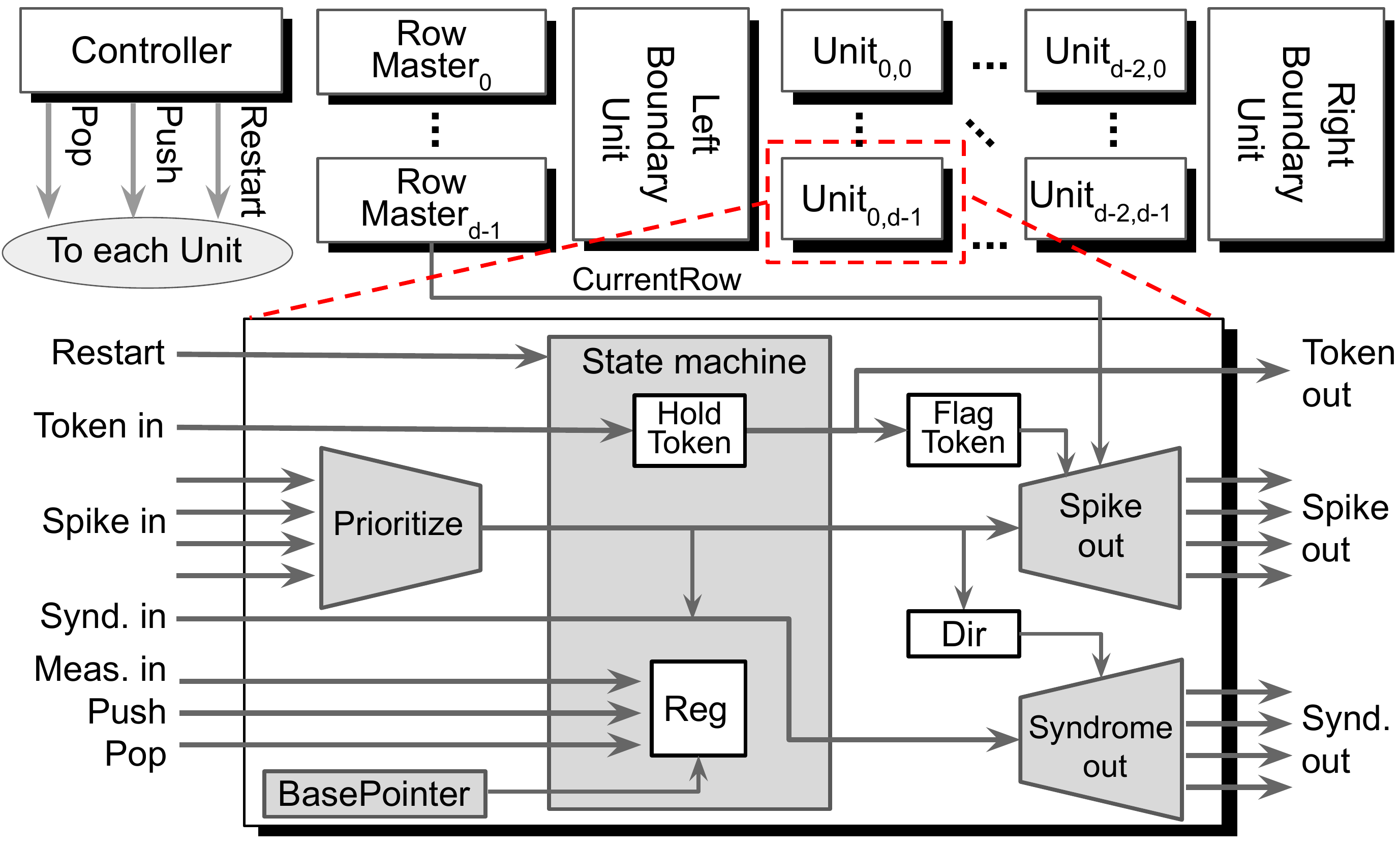}
    \caption{Overview of the architecture and the microarchitecture of Unit}
    \label{fig:arch_and_datapath}
    \vspace{-2mm}
\end{figure}

Figure~\ref{fig:arch_and_datapath} shows the overview of the hardware architecture for $X$ error detection for a single distance-$d$ logical qubit\footnote{The identical hardware applies to $Z$ error detection.}. 
Each hardware \textit{Unit} corresponds to the ``Unit'' of Algorithm~1, and $d \times (d-1)$ Units are aligned in a 2-D grid pattern. There is one Controller per one logical qubit to orchestrate the Units by distributing \textit{Push}, \textit{Pop}, and \textit{Restart} signal. 
Note that this architecture can be easily extended to any code distance $d$ thanks to its distributed nature.

Each Unit has a register (Reg) that stores measured values of the corresponding ancilla bit. It works as a queue;
whenever the measurement process is performed, the Controller sends the Push signal to all the Units and the measured value is stored at the end of Reg.
When the first bit entry of Reg (corresponding to the oldest measurement) of all the Units becomes 0, the error correction process for the current layer is completed. In this case, the Controller broadcasts the Pop signal, which makes the values in the Reg being one-bit shifted.
Based on the results of Section~\ref{subsec:pre_error_correction_performance}, we assume $th_{v} = 3$ which means the Reg needs to store at least three measurement values. In this paper, we set the size of a Reg to 7-bit with some margin.

A ``Row Master'' module is attached to each row of the 2-D Unit array and gives the Token to the first Unit of the row.
It always checks Reg values in all the Units in the row, and if none of the bits in Reg is 1, it avoids giving the Token to the row to avoid meaningless Token passing on non-erroneous Units which helps reduce the time taken for QEC. 
In this case, it just passes the Token to the next Row Master. 
It is also responsible to send \textit{CurrentRow} signal to all the Units of the row. 

Two \textit{Boundary Units} are located on each side of the Unit array. 
Each of them is connected to $d$ Units on a horizontal edge. 
One Boundary Unit can be shared by all the Units on each edge. 
Distributing a Spike to all the Units does not affect QEC's correctness since only the first Spike that reaches the Sink Unit is valid in the algorithm.

\subsection{Unit implementation\label{subsec:neuron_unit}}

In each Unit, there are five components as described below.

\begin{itemize}
    \item \textbf{State machine}: A state machine controls the behavior of the Unit. The state transition happens based on Restart signal, coming Token, and coming Spike. This module has two register memories; HoldToken and Reg, and their values affect state transition.

    \item \textbf{Prioritization module}: As a Unit may get multiple Spikes from different directions simultaneously, we need to prioritize the input from a specific direction. This module selects one Spike from them by the predefined priority. We use the race logic concept, which utilizes the relative propagation time of signals. We put an appropriate signal delay in each direction for coming Spike so that a Spike from the highest priority direction must come faster than the others. 

    \item \textbf{Spike out module}: This module sends a Spike to the appropriate direction based on values of the CurrentRow and the FlagToken.

    \item \textbf{Syndrome out module}: This module sends a Syndrome signal to the direction indicated by the \textit{Dir} register that stores the opposite direction of the coming Spike. It also generates a correction signal to the associated data qubit.

    \item \textbf{BasePointer module}: This module controls which bit position in Reg to be read out based on the value of the base register. 
    The read value from Reg is used to decide whether the Unit should send a Spike. 
\end{itemize}

\subsection{SFQ logic gates}

\begin{table}[t]
\centering
\caption{Summary of SFQ logic elements\label{tab:SFQ_logic_elements}}
\vspace{-2mm}
\scalebox{0.9}{
\begin{tabular}{|l|c|c|c|c|} \hline
cell                          & JJs &\begin{tabular}[c]{@{}c@{}}Bias current\\ (mA)\end{tabular} & \begin{tabular}[c]{@{}c@{}}Area\\ ($\mu \mathrm{m}^2$)\end{tabular}& \begin{tabular}[c]{@{}c@{}} Latency\\ (ps)\end{tabular}\\ \hline \hline
splitter                      & 3        & 0.300 & 900   & 4.3   \\
merger                        & 7        & 0.880 & 900   & 8.2   \\
1:2 switch                    & 33       & 3.464 & 8100  & 10.5    \\
destructive readout (DRO)     & 6        & 0.720 & 900   & 5.1   \\
nondestructive readout (NDRO) & 11       & 1.112 & 1800  & 6.4    \\
resettable DRO (RD)           & 11       & 0.900 & 1800  & 6.0 \\ 
dual-output DRO (D2)          & 12       & 0.944 & 1800  & 6.8   \\ \hline 
%AND                           & 14       & 1.428      \\  \hline
\end{tabular}
}
\vspace{-4mm}
\end{table}

We designed the QECOOL hardware based on an RSFQ cell library \cite{detail_of_cell_library_ADP2} developed for a niobium nine-layer, 1.0-$\mu$m fabrication technology \cite{Shuichi_NAGASAWA2014,Akira_FUJIMAKI2014}.
Table~\ref{tab:SFQ_logic_elements} summarizes the SFQ logic gates used in this work.
Since the essential element of SFQ that affects power consumption and hardware cost is Josephson junction (JJ), 
the Table shows the number of JJs for each gate and assumed the bias current required for operation.
The operating temperature and designed supply voltage are 4-K and 2.5~mV, respectively. 
In RSFQ, since most of the power is consumed statically almost independent of switching activities, it is calculated by multiplying the bias voltage and currents.

\begin{table}[tbh]
\centering
\caption{The total number of logic elements, the number of JJs, circuit area and the of each module that constitutes an ancilla unit based on the AIST 10-kA/cm$^2$ ADP cell library\cite{detail_of_cell_library_ADP2}.}
\label{tab:JJcounts_and_power consumption}
\vspace{-2mm}
\scalebox{0.8}{
\begin{tabular}{|l|c|c|c|c|c|c||c|} \hline
cell       & \begin{tabular}[c]{@{}l@{}}State\\  machine\end{tabular} & \begin{tabular}[c]{@{}l@{}}Prio-\\ritiz-\\ ation\end{tabular} & \begin{tabular}[c]{@{}l@{}}Base \\ pointer \\ (7-bit)\end{tabular} & \begin{tabular}[c]{@{}l@{}}Spike\\ out\end{tabular} & \begin{tabular}[c]{@{}l@{}}Synd-\\rome\\ out\end{tabular} & Other & \begin{tabular}[c]{@{}l@{}}Total\\ (7-bit)\end{tabular}  \\ \hline
splitter              & 17     & 4    &        & 8     &     & 2     & 31        \\
merger                & 14     & 9    & 30     & 8     & 2      & 2     & 65         \\
1:2 switch            & 8      &       &        & 3      &       &       & 11         \\
DRO                   &        &       & 3      &        &       &       & 3          \\
NDRO                  &        &       & 20     &        &       &       & 20         \\
RD                    & 6      &       & 30     & 4      & 4     &       & 44         \\ 
D2                    &        &       & 6      &        &       &       & 6         \\ 
Wire                  & 196    & 82    & 1085   & 91     &      & 18    & 1472         \\ \hline \hline
Total JJs             & 675    & 157   & 1935   & 314    & 58    & 38    & 3177    \\
\begin{tabular}[c]{@{}l@{}}Total area\\ ($\mu m^2$)\end{tabular}            & 265500 & 82800 & 709200 & 129600 & 25200 & 62100 & 1274400 \\
\begin{tabular}[c]{@{}l@{}}Total bias \\ current (mA)\end{tabular}  & 69.7    & 15.3   & 208.5   & 32.2    & 5.4    & 5.0    & 336 \\ 
Latency (ps)        & 98.7       & 28.0      & 147     & 61.1    & 10.4     &     & 215   \\\hline  %TODO 11/19

\end{tabular}
}
\end{table}

\begin{figure}[t]
    \centering
    \includegraphics[width=0.85\linewidth]{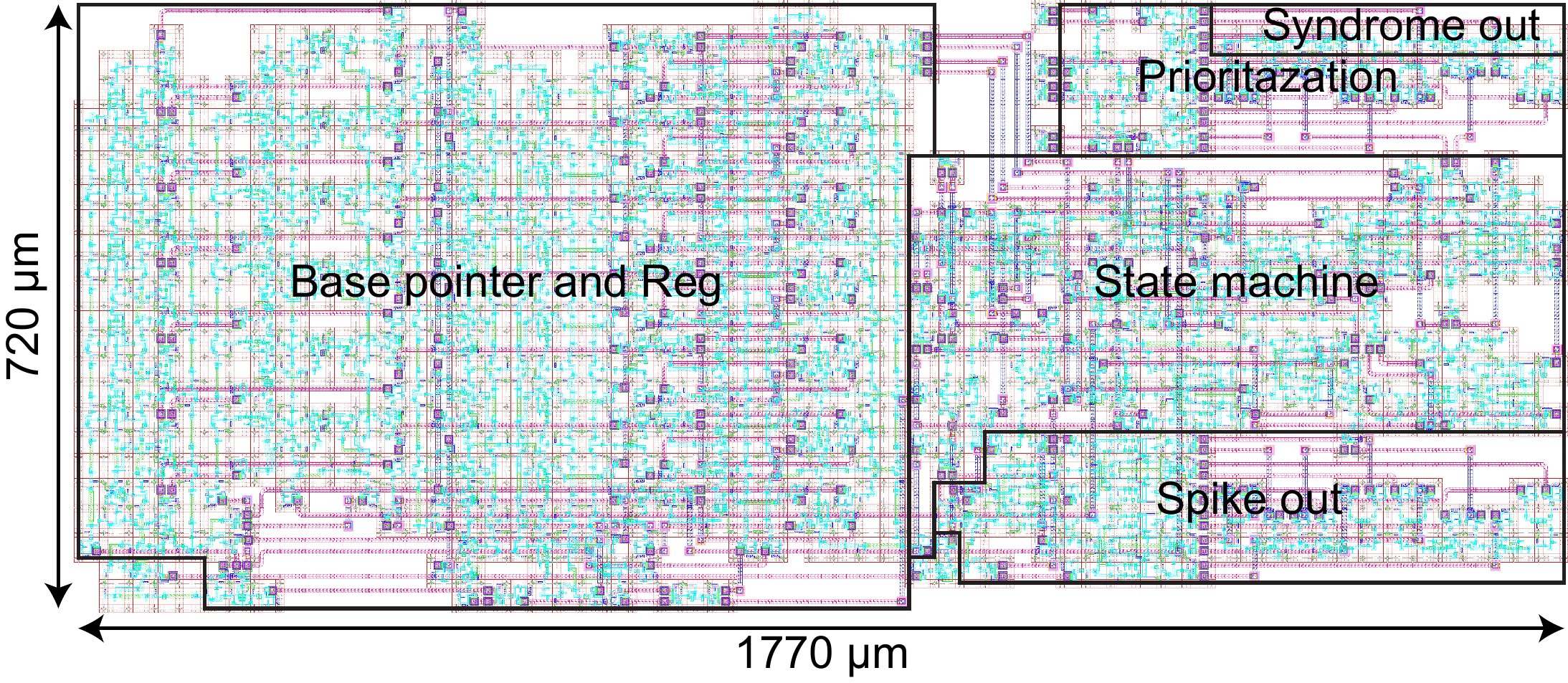}
    \caption{The layout of the designed QECOOL Unit}
    \label{fig:layout}
\end{figure}

We used the Josephson simulator (JSIM)\cite{jsim}, a SPICE-level simulator, to verify the functionality of the designed Unit with 7-bit Reg and evaluate its latency.
Table~\ref{tab:JJcounts_and_power consumption} shows the total number of JJs, total area, total bias current, and latency of each module.
Figure~\ref{fig:layout} shows the layout of the QECOOL Unit.
A Unit consists of 3177~JJs in total, and its area footprint is $1.274~\mathrm{mm}^2$.
The maximum delay of the designed circuit is 215~ps, results in the maximum operating frequency of about 5~GHz. 
It can operate fast enough to achieve the required QEC latency described later in Section~\ref{subsec:execution_cycles}.
The power consumption of a Unit is $336_{\texttt{[mA]}} \times 2.5_{\texttt{[mV]}} = 840_{[\mu\texttt{W}]}$ including the wiring power, if implemented with RSFQ logic.

%\section{Evaluations\footnotemark \label{sec:experiment}}
\section{Evaluations of the online-QEC decoder}

\subsection{Execution cycles\label{subsec:execution_cycles}}

\begin{table}[tbh]
\centering
\caption{Per layer execution cycles of QECOOL} 
\label{tab:execution_cycle}
\vspace{-2mm}
\scalebox{0.9}{
\begin{tabular}{|c|c|c|c|c|c|c|c|c|c|} \hline
   & \multicolumn{3}{c|}{$p = 0.001$} & \multicolumn{3}{c|}{$p = 0.005$}  & \multicolumn{3}{c|}{$p = 0.01$}   \\ \hline
   $d$& Max   & Avg     & $\sigma$  & Max   & Avg     &$\sigma$& Max & Avg & $\sigma$  \\ \hline \hline
5  & 104   & 6.10 & 4.99 & 144   & 10.4  & 11.2  & 166  & 15.6  & 15.8  \\
7  & 303   & 11.8 & 14.5 & 515   & 28.7  & 30.1  & 557  & 47.4  & 43.9  \\
9  & 800   & 22.7 & 30.6 & 1018  & 64.2  & 57.7  & 1308 & 107 & 89.7  \\
11 & 996   & 41.6 & 53.6 & 1779  & 120   & 95.3  & 2435 & 201 & 161 \\
13 & 1890  & 71.3 & 82.9 & 3289  & 199   & 147   & 4072 & 337 & 266 \\ \hline
\end{tabular}
}
\vspace{-1mm}
\end{table}

Table~\ref{tab:execution_cycle} shows the number of execution cycles per layer for QECOOL for several combinations of coding distance $d$ and physical error rate $p$. 
It is observed that the execution cycles of QECOOL are highly dependent on $d$ and $p$. 

Since it is reported that measuring ancilla bit takes about $1 \mu\mathrm{s}$ \cite{Gidney2019HowTF}, QEC is fast enough if the process for one layer is finished within $1 \mu\mathrm{s}$.

\subsection{Error correction performance\label{subsec:error_correction_performance}}
%In this subsection, we evaluate the error correction performance of the proposed algorithm when executed in a real-time processing manner.
We use the same error model as in Subsection \ref{subsec:pre_error_correction_performance} to evaluate the performance of QECOOL for online-QEC.
The measurement process is assumed to be performed once every 1~$\mu$s.
Each Unit has 7-bit Reg, and $th_{v}$ in Algorithm~1 is set to 3.
If Reg overflows because of the slow QEC performance, the trial is considered as a failure.

\begin{figure}[t]
    \centering
    \includegraphics[width=0.85\linewidth]{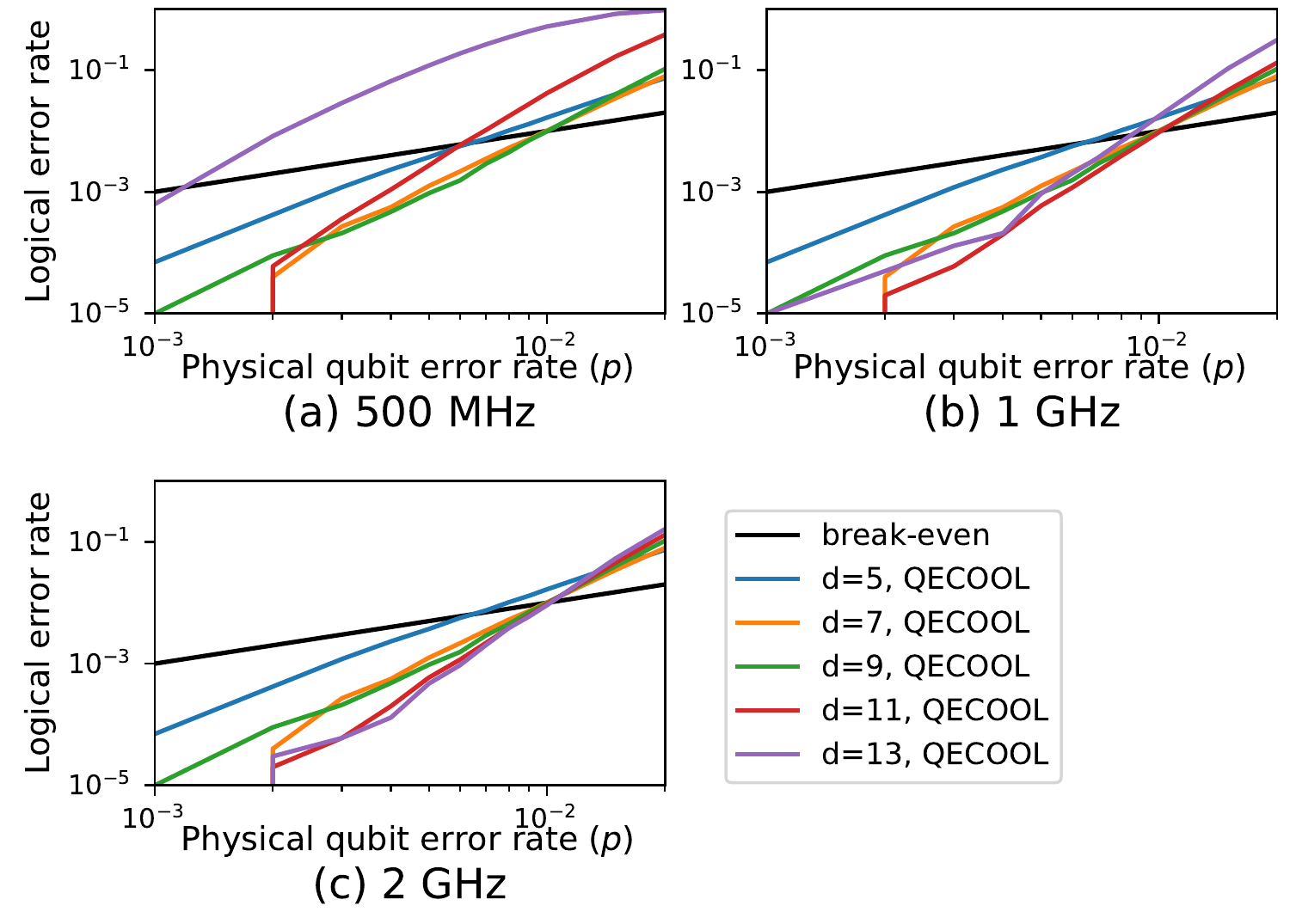}
    \vspace{-1mm}
    \caption{Physical vs. logical error rate plot for the QECOOL algorithm operating at several frequencies. }
    \label{fig:error_correction_performance_realtime}
    \vspace{-2mm}
\end{figure}

Figures~\ref{fig:error_correction_performance_realtime}(a) and (b) indicate that the slower frequency causes buffer overflow of Unit, which affects the error correction performance at larger code distance $d$.
Only in Figure~\ref{fig:error_correction_performance_realtime}(c), we can observe the $p_{th}$ of QECOOL at approximately $p = 1.0\%$, which is slightly smaller than that of batch-QECOOL. 

\subsection{Power estimation with ERSFQ logic}
To put the more decoder units in a dilution refrigerator, their power consumption should be much lower, and the RSFQ technology is not feasible. Instead, we need to use the ERSFQ\cite{ERSFQ} technology where the static power consumption is eliminated. Power consumption in ERSFQ circuits is only consumed by dynamic power twice as high as that of RSFQ.
%In an ERSFQ design, the bias supply resistors of a conventional RSFQ circuit are replaced by JJs while the gate configuration is the same as that of RSFQ.
Though ERSFQ is slower than RSFQ, this is not a problem for our hardware design since our target clock frequency is much lower than the maximum frequency in RSFQ.

Based on the RSFQ design and the power model of ERSFQ\cite{ERSFQ_model}, we estimate the power consumption when ERSFQ is applied.
Here, the power of a Unit with ERSFQ can be estimated as follows:
\begin{flushleft}
$P_{unit} =(\text{bias current}) \times (\text{frequency}) \times \Phi_{0} \times 2$
\end{flushleft}
We use flux quantum $\Phi_0$ of $2.068 \times 10^{-15}$~Wb. The reason why it is multiplied by 2 is to represent twice of dynamic power in ERSFQ.

For a proposed Unit with 7-bit Reg, the total bias current is 336~mA. 
If we suppose 2~GHz clock frequency, the power consumption of a Unit at a 4-K environment is estimated as follows:
\begin{eqnarray} \nonumber
    336_{[\text{mA}]} \times 2_{[\text{GHz}]} \times (2.068 \times 10^{-15})_{[\text{Wb}]} \times 2 
    = 2.78_{[\text{$\mu$W / Unit}]}.
\end{eqnarray}

%\begin{eqnarray} \nonumber
%    (336 \times 10^{-3})_{[\text{A}]} \times (2.0 \times 10^9)_{[\text{Hz}]} \times (2.068 \times 10^{-15})_{[\text{Wb}]} \times 2 \\ \nonumber
%    = 2.78_{[\text{$\mu$W / Unit}]}.
%\end{eqnarray}
%Suppose that the cooling capacity of a cryocooler is about $1~W$ at 4~K\cite{cryogenic_quantum}, around 360,000 Units can be co-located. 
%A logical qubit encoded with a distance-$d$ SC requires $2\times d(d-1)$ Units because it has $d(d-1)$ X and Z ancilla bits, respectively.
%Hence our implementation would protect around 300 logical qubits of code distance $d = 27$, or over 3,000 logical qubits of code distance $d = 9$.

\subsection{Comparison to existing decoders}

\begin{table}[t]
\centering
\caption{Qualitative comparison of decoder performance \label{tab:decoder_comparison}}
\vspace{-2mm}
\begin{tabular}{|c|c|c|c|} \hline
Decoder         & \begin{tabular}[c]{@{}c@{}}$P_{th}$ (2-D / 3-D)\end{tabular} & Latency & Environment        \\ \hline \hline
MWPM\cite{MWPM_for_surface_code}            & 10.3\% / 2.9\%&  High  & Software           \\ \hline
UF\cite{almost_linear}      & 9.9\% / 2.6\% &  Medium  & FPGA \cite{das2020scalable}       \\ \hline
AQEC\cite{holmes2020nisq}            & 5\% / -  & Very low    & SFQ\\ \hline
QECOOL & 6.0\% / 1.0\%   & Low & SFQ\\ \hline
\end{tabular}
\vspace{-3mm}
\end{table}

Table~\ref{tab:decoder_comparison} shows a brief comparison of MWPM and recent prominent hardware-efficient decoding algorithms.
While MWPM and union-find (UF) have a higher threshold, they are designed to operate in a room-temperature environment as they are implemented by software or an FPGA device.
%For qubits built on superconducting circuits operating in a cryogenic environment, it is necessary to take into account the delay and power consumption due to interconnection between different environments.

\begin{table}[t]
\centering
\caption{Comparison of AQEC\cite{holmes2020nisq} and QECOOL \label{tab:aqec_comparison}}
\vspace{-2mm}
\scalebox{0.81}{
\begin{tabular}{|l||l|l|l|l|l|l|} \hline
         & \begin{tabular}[c]{@{}l@{}}$p_{th}$\\ (2-D/3-D)\end{tabular} & \begin{tabular}[c]{@{}l@{}}Execution \\ time \\ per layer\\ (Max/Avg.)\\ ($ns$)\end{tabular} & \begin{tabular}[c]{@{}l@{}}Power \\ per \\ Unit\\ ($\mu W$)\end{tabular} & \begin{tabular}[c]{@{}l@{}}\# Units \\ required \\ per \\ logical\\ qubit\end{tabular} & \begin{tabular}[c]{@{}l@{}}Directly \\appli-\\ cable \\ to 3-D\end{tabular}    & \begin{tabular}[c]{@{}l@{}}\# protect-\\ able \\ logical \\ qubits\end{tabular} \\ \hline \hline

AQEC     & 5.0\% / -                                                                        & 19.8 / 3.93                                                                          & 13.44                                                                                       & $(2d-1)^2$                                                                                   & No   & 37                                                                                                            \\\hline
\begin{tabular}[c]{@{}l@{}}QECOOL\\ (7-bit Reg)\end{tabular} & \begin{tabular}[c]{@{}l@{}}6.0\% /\\1.0\% \end{tabular}                                                                      & 400 / 20.8                                                                          & 2.78                                                                                        & $2d(d-1)$                                                                                    & Yes                                                                & 2498 \\    \hline                                          
\end{tabular}
}
\vspace{-3.5mm}
\end{table}
% AQEC     & 5.0\% / - \footnotemark                                                                       & 19.8 / 3.93  %\footnotetext{Since threshold values are reduced by 70-80\% for any method in Table \ref{tab:decoder_comparison} by extending from 2-D to 3-D, we expect $p_{th}$ of 3-D AQEC to be slightly smaller than ours.} 

AQEC\cite{holmes2020nisq} and our decoder are designed to operate in a cryogenic environment and have higher scalability than others, though these have a slightly lower threshold. 
We summarize a detailed comparison of AQEC and QECOOL in Table~\ref{tab:aqec_comparison}.
We assume the coding distance $d$ of 9, the longest one evaluated in the AQEC paper, and the physical error rate (for both data and ancilla qubits) $p$ of 0.001, which is 0.1 times the $p_{th}$ of our decoder.
The power budget of the 4-K temperature region of dilution refrigerators is supposed to be 1~W\cite{cryogenic_quantum}, and the number of protectable qubits is estimated in terms of the decoders' power consumption.
Applying the same arguments as Subsection~\ref{subsec:pre_error_correction_performance}, we assume that extending AQEC to 3-D requires 7 times the modules needed for 2-D processing.
In addition, we optimistically consider AQEC's latency for 2-D processing as its per layer latency when extended to 3-D.
The Table shows that QECOOL is superior to AQEC in other aspects than latency, which is not a major disadvantage since our decoder can finish one layer process within 1 $\mu$s, the interval of the measurement process\cite{Gidney2019HowTF}.
AQEC can lower its operating frequency to reduce its power consumption by taking advantage of its low latency, however, our method still has lower power consumption, even taking this into account.
Note that the $p_{th}$ of AQEC for the 3-D case is unknown, and it is expected to be slightly smaller than ours as
the $p_{th}$ trends of 2-D to 3-D in Table~\ref{tab:decoder_comparison} indicate 70-80\% reduction in 3-D cases.

%Our QEC method has higher threshold and capability of dealing with the 3D surface code compared with AQEC. 
%This is the great advantage of the proposed hardware implementation. 

%\addtocounter{footnote}{1}

%\addtocounter{footnote}{1}
 %\footnotetext{We consider the latency for 2-D processing as its per layer latency when extended to 3-D, which is an optimistic estimation, and it is expected to be larger in practice due to the increase in complexity of the process.}
%\addtocounter{footnote}{1}
% \footnotetext{Applying the same arguments as Subsection\ref{subsec:pre_error_correction_performance}, we assume that extending AQEC to 3-D requires 7 times the modules needed for 2-D processing.}

\section{Conclusion}
\label{sec:conclusion}
% We focused on efficient quantum error correction (QEC) using SFQ circuits.
In this paper, we proposed an online-QEC algorithm named QECOOL for decoding surface code with measurement errors and designed a decoder with SFQ logic based on the algorithm. 
We evaluated the circuit characteristics and the error-correcting performance of our design.
As a result, our decoder is power-efficient enough to protect around 2500 logical qubits with distance-9 surface code in the 4-K layer of a dilution refrigerator, and it is fast and accurate enough for online-QEC.

\bibliographystyle{abbrv}
\bibliography{references}

\end{document}